\documentclass[10pt,american,english,aps,showkeys,showpacs,nofootinbib]{revtex4-1}
\usepackage[T1]{fontenc}
\usepackage{textcomp}
\usepackage[utf8]{inputenc}
\usepackage{babel}
\usepackage{amsmath}
\usepackage{amssymb}
\usepackage{stackrel}
\usepackage{graphicx}
\usepackage{geometry}
\usepackage[bookmarks=false,
 breaklinks=false,pdfborder={0 0 1},backref=section,colorlinks=false]
 {hyperref}

\makeatletter
\usepackage{babel}
\usepackage{todonotes}
\usepackage[bottom]{footmisc}
\usepackage{graphicx}
\usepackage{dcolumn}
\usepackage{bm}
\usepackage{todonotes}
\makeatother

\begin{document}
\title{Dirac Oscillator for Spin-1/2 Particles in a Spinning Cosmic String
Spacetime with Spacelike Disclination and Dislocation}
\author{Abdelmalek Boumali}
\email{boumali.abdelmalek@gmail.com}

\affiliation{Laboratory of theoretical and applied Physics~\\
 Echahid Cheikh Larbi Tebessi University, Algeria}
\date{\today}
\begin{abstract}
We solve the covariant Dirac-oscillator problem for spin-½ particles
in the spacetime of a spinning cosmic string endowed with both a conical
disclination and a screw-type dislocation. Working in a tetrad basis,
we reduce the Dirac equation to a single radial equation and map it
to the confluent-hypergeometric form, enabling exact normalizable
solutions and a closed quantization rule. The resulting spectrum is
implicit in the energy and exhibits a defect-renormalized angular
index that depends on curvature (deficit parameter $\alpha$) and
on two torsional moments $J_{t}$ (time-like/spin) and $J_{z}$ (space-like/screw),
thereby coupling geometry, spin, and longitudinal momentum $k$. We
analyze three limiting configurations balanced torsion $J_{t}=J_{z}$,
purely spinning $J_{z}=0$, and purely screw $J_{t}=0$ and the fully
coupled case, showing how time-like torsion induces an $E$-dependent,
self-consistent shift, whereas space-like torsion contributes an explicit
$k$-dependent bias, breaking flat-space degeneracies and distorting
$\ell$-resolved level spacings. Positivity of $g_{\phi\phi}$ imposes
a minimal radius when $|J_{t}|>|J_{z}|$, which acts as a geometric
hard wall and further enhances the effective centrifugal barrier.
All special limits ($J_{t}\!\to\!0$, $J_{z}\!\to\!0$, $\alpha\!\to\!1$)
recover the appropriate sub-cases and, ultimately, the Moshinsky result.
The model clarifies spin--gravity--momentum interplay in torsion-rich
backgrounds and suggests testable analogs in Dirac materials and quantum
simulators.
\end{abstract}
\maketitle

\section{Introduction }

The Dirac oscillator (DO) stands as a fundamental model in relativistic
quantum mechanics, notable for its exact solvability and wide-ranging
applicability across theoretical and applied physics. Originally proposed
by Itô, Mori, and Carrière \cite{ito1967}, the model modifies the
standard Dirac equation through the substitution $\mathbf{p}\rightarrow\mathbf{p}-im\omega\beta\mathbf{r}$,
where $m$ is the mass of the particle, $\omega$ denotes the oscillator
frequency, and $\beta$ is the Dirac matrix. This prescription yields
a relativistic system whose non-relativistic limit reduces to a harmonic
oscillator augmented by strong spin--orbit coupling \cite{moshinsky1989,moreno1989,benitez1990}.
Owing to these features, the Dirac oscillator has found applications
in diverse fields including nuclear physics, quantum optics, and condensed
matter theory, where it serves as a powerful framework for analyzing
relativistic bound states under various physical conditions \cite{boumali2013,boumali2015,Franco-Villafane2013,BakkeFurtado2009,quimbay2013a}.

At the same time, significant efforts have been directed toward understanding
how topological defects influence quantum fields. These spacetime
defects---such as cosmic strings---arise naturally from symmetry-breaking
phase transitions in the early universe, leading to geometric features
like curvature (associated with disclinations) and torsion (associated
with dislocations). Such geometric alterations profoundly impact quantum
systems by modifying their boundary conditions, symmetry structures,
and energy spectra. Earlier studies explored the hydrogen atom within
spacetimes containing cosmic string or monopole backgrounds, and this
line of inquiry was later extended to scalar and spinor oscillators
in curved spacetimes, incorporating magnetic fields, torsion, and
dislocations. Additional related investigations can also be found
throughout the literature case\cite{BakkeFurtado2009,BoumaliMessai2014,MessaiBoumali2015,BoumaliMessai2017,ChenLongRanYangLong2020,MarquesEtAl2005,StrangeRyder2016,HosseinpourHassanabadideMontigny2019,Oliveira2019,CunhaDiasSilva2020,Bakke2013,Carvalho2011,MotaBakke2017,Castro2016,HosseinpourHassanabadi2015,SantosBarros2018,BakkeFurtado2010}.

More broadly, scalar and spinor fields have been explored in geometries
with diverse topological features \cite{VilenkinShellard1994}. Investigations
include exact solutions for scalar particles in spinning cosmic string
backgrounds, studies of non-inertial effects on the Dirac oscillator
\cite{HosseinpourHassanabadi2015,HosseinpourHassanabadideMontigny2019,CunhaDiasSilva2020}.
Furthermore, the influence of torsion on fermionic dynamics has been
investigated in both gravitational and condensed matter analogs, where
disclinations and screw dislocations effectively mimic gravitational
phenomena \cite{Figielski2002,MarquesEtAl2005}. From a cosmological
perspective, topological defects such as cosmic strings also have
astrophysical implications, including gravitational lensing and perturbations
in density distributions, with scattering cross-sections modulated
by the Fourier transform of the density correlation function \cite{BakkeFurtado2009,BakkeFurtado2010}.

Motivated by these considerations, the present work investigates the
Dirac oscillator for spin-½ particles in the spacetime of a spinning
cosmic string endowed with both a spacelike disclination (curvature)
and a screw-type dislocation (torsion). The geometry under study features
non-diagonal metric components and is formulated using a local tetrad
basis in cylindrical coordinates. The covariant Dirac equation is
then derived and exactly solved by isolating the lower spinor component,
leading to a second-order differential equation that governs the system's
radial dynamics.

\selectlanguage{american}%
Our contribution is not the use of a Dirac equation in a fixed background,
but an exact, closed-form treatment of the Dirac oscillator in a torsion-rich
spinning--string geometry that combines disclination ($\alpha<1$)
with both time-like ($J_{t}$) and space-like ($J_{z}$) torsion.
Within a single framework we analyze the balanced $(J_{t}=J_{z})$,
purely spinning $(J_{z}=0)$, purely screw $(J_{t}=0)$, and fully
coupled cases, obtaining spectra that (i) exhibit defect-renormalized
angular indices, (ii) display degeneracy breaking across $\ell$,
and (iii) show momentum-dependent corrections via $k$ that are absent
in flat space. Because the radial problem maps exactly to the confluent--hypergeometric
form, these effects are derived analytically, allowing us to isolate
which features arise from curvature versus torsion. This positions
the model as a clean benchmark for spin--geometry--momentum coupling
and for analog implementations (graphene-like and photonic/ion platforms)
discussed later.

\selectlanguage{english}%
Our results reveal that both the curvature, encapsulated by the angular
deficit parameter $\alpha$, and the torsion, characterized by parameters
$J_{t}$ and $J_{z}$, have pronounced effects on the relativistic
energy spectrum. In particular, they introduce effective shifts in
the angular quantum number, resulting in the lifting of degeneracies
and the emergence of energy- and momentum-dependent spectral deformations.
These effects signify a nontrivial coupling between spin, angular
momentum, and the background geometry, offering a new perspective
on spin--gravity interactions in torsion-rich spacetimes. Each physical
configuration---balanced torsion, purely temporal torsion, and purely
spatial torsion---exhibits distinct spectral behavior, highlighting
the intricate interplay between geometry and relativistic quantum
dynamics.

This exactly solvable model contributes to the broader understanding
of spinor fields in non-Euclidean geometries and may serve as a theoretical
framework for modeling analogous effects in condensed matter systems,
such as those found in graphene or cold-atom simulations of torsional
geometries. Future extensions could include the incorporation of external
electromagnetic fields, thermal fluctuations, or supersymmetric modifications,
further enriching the connection between quantum field theory and
geometrically nontrivial backgrounds.

The structure of the paper is organized as follows. In Section II,
we present the spinning cosmic string spacetime, construct the appropriate
tetrads, and derive the covariant Dirac equation incorporating the
oscillator interaction. In Section III, we analyze three physically
significant configurations: (i) balanced torsion, where temporal and
spatial contributions are equal; (ii) purely temporal torsion, corresponding
to a spinning string; and (iii) purely spatial torsion, associated
with screw dislocations. For each case, we derive exact energy spectra
and wavefunctions, and compare them to the flat-space Moshinsky oscillator.
In Section IV, we summarize our main findings, discuss their physical
implications, and suggest directions for future research, particularly
in extending the model to more general spacetimes or interacting quantum
systems.
\selectlanguage{american}%

\section{DIRAC OSCILLATOR IN THE COSMIC STRING BACKGROUND}

\selectlanguage{english}%
In this section, we address the solution of the Dirac oscillator in
the presence of a cosmic string background characterized by the spacetime
signature $(-+++)$. The geometry induced by the cosmic string is
described using cylindrical coordinates $(t,\rho,\phi,z)$, and the
corresponding line element for a straight, rotating cosmic string
endowed with torsion is given by \cite{PuntigamSoleng1997,Ozdemir2005,Jusufi2016}:
\begin{equation}
ds^{2}=-\left(dt+4GJ^{t}d\varphi\right)^{2}+d\rho^{2}+\alpha^{2}\rho^{2}d\varphi^{2}+\left(dz+4GJ^{z}d\varphi\right)^{2},\label{1}
\end{equation}
where the parameter $\alpha$, satisfying $0<\alpha<1$, represents
the angular deficit associated with the conical geometry. The quantity
$J^{t}$ denotes the linear density of angular momentum, responsible
for frame-dragging effects, while $J^{z}$ characterizes the screw-dislocation
parameter linked to torsion. 

The coordinate ranges are defined as $-\infty<t,z<\infty$, $0\leq\rho<\infty$,
and $0\leq\phi\leq2\pi$. The deficit parameter \$\textbackslash alpha\$
is associated with the conical structure of the spacetime and satisfies
the relation $\alpha=1-4\mu$, where \$\textbackslash mu\$ is the
linear mass density of the cosmic string expressed in natural units.

\selectlanguage{american}%
The metric tensor $g_{\mu\nu}$ in matrix form is:
\begin{equation}
g_{\mu\nu}=\begin{pmatrix}-1 & 0 & -4GJ^{t} & 0\\
0 & 1 & 0 & 0\\
-4GJ^{t} & 0 & \alpha^{2}\rho^{2}-16G^{2}[(J^{t})^{2}-(J^{z})^{2}] & 4GJ^{z}\\
0 & 0 & 4GJ^{z} & 1
\end{pmatrix}\label{2}
\end{equation}
We note here that for the metric in Eq. (\ref{1}), the azimuthal
component is $g_{\phi\phi}=\alpha^{2}\rho^{2}-16G^{2}[(J_{t})^{2}-(J_{z})^{2}]$.
Physical admissibility requires $g_{\phi\phi}>0$ to ensure a positive
definite metric signature and avoid closed timelike curves or unphysical
regions\cite{Vitoria2025_IJMPA_2450176,Bakke2020_IJMPA_2050129}.
Hence, when $(J_{t})^{2}>(J_{z})^{2}$, we restrict the configuration
space to 
\begin{equation}
\rho>\rho_{c}=\frac{4G}{\alpha}\sqrt{(J_{t})^{2}-(J_{z})^{2}}.\label{2-1}
\end{equation}
For $J_{z}=0$, this reduces to $\rho>4G|J_{t}|/\alpha$; for $J_{t}=0$,
the inequality is automatically satisfied for all $\rho>0$. 

This radial cutoff $\rho_{c}$ has important implications for the
quantum system. All normalization integrals for the wavefunctions
are thus evaluated over $\rho\in(\rho_{c},\infty)$ (or $(0,\infty)$
when $\rho_{c}=0$), ensuring that the probability density is confined
to the physically admissible region. Boundary conditions at $\rho=\rho_{c}$
are imposed such that the wavefunction vanishes or satisfies the conditions
that prevent leakage into the forbidden zone, analogous to hard-wall
potentials in defect spacetimes. This restriction modifies the effective
centrifugal barrier in the radial equation, potentially shifting the
energy levels and altering the density of states. In particular, for
large torsion parameters, $\rho_{c}$ introduces a minimal radius
that lifts low-angular-momentum degeneracies and affects the ground-state
energy, providing a geometric regularization akin to those in rotating
frames \cite{Vitoria2025_IJMPA_2450176,Bakke2020_IJMPA_2050129}.

\selectlanguage{english}%
Now, the physical properties of this spacetime are determined by the
value of $j^{2}=(J^{t})^{2}-(J^{z})^{2}$ \cite{Ozdemir2005}: 
\begin{itemize}
\item Case (1): $j^{2}=0$ (i.e., $|J^{z}|=|J^{t}|$)
\begin{equation}
ds^{2}=-\left(dt+4GJ\,d\varphi\right)^{2}+d\rho^{2}+\alpha^{2}\rho^{2}d\phi^{2}+\left(dz+4GJ\,d\varphi\right)^{2}\label{3}
\end{equation}
with
\begin{equation}
g_{\mu\nu}=\begin{pmatrix}-1 & 0 & -4GJ^{t} & 0\\
0 & 1 & 0 & 0\\
-4GJ^{t} & 0 & \alpha^{2}\rho^{2} & 4GJ^{z}\\
0 & 0 & 4GJ^{z} & 1
\end{pmatrix}\label{4}
\end{equation}
This describes a string interacting with a circularly polarized plane-fronted
gravitational wave.
\item Case (2): $|J^{z}|=0$
\begin{equation}
ds^{2}=-\left(dt+4GJ^{t}d\varphi\right)^{2}+d\rho^{2}+\alpha^{2}\rho^{2}d\varphi^{2}+dz^{2}\label{eq:5}
\end{equation}
with
\begin{equation}
g_{\mu\nu}=\begin{pmatrix}-1 & 0 & -4GJ^{t} & 0\\
0 & 1 & 0 & 0\\
-4GJ^{t} & 0 & \alpha^{2}\rho^{2}-16G^{2}(J^{t})^{2} & 0\\
0 & 0 & 0 & 1
\end{pmatrix}\label{6}
\end{equation}
This corresponds to a spinning cosmic string without any spatial dislocation.
\item Case (3): $|J^{t}|=0$
\begin{equation}
ds^{2}=-dt^{2}+dr^{2}+\alpha^{2}\rho^{2}d\varphi^{2}+\left(dz+4GJ^{z}d\varphi\right)^{2}\label{7}
\end{equation}
with
\begin{equation}
g_{\mu\nu}=\begin{pmatrix}-1 & 0 & 0 & 0\\
0 & 1 & 0 & 0\\
0 & 0 & \alpha^{2}\rho^{2}+16G^{2}(J^{z})^{2} & 4GJ^{z}\\
0 & 0 & 4GJ^{z} & 1
\end{pmatrix}\label{8}
\end{equation}
In this scenario, the spacetime describes screw dislocations, which
can be interpreted as a combination of a screw dislocation (with $2GJ^{z}/\pi$
analogous to a Burgers vector) and a disclination
\end{itemize}
The governing equation for the spinor field in this curved background
is the Dirac equation \cite{ObukhovHehl2004,Pollock2010,Fulling1989,Nakahara2003,ParkerToms2009,PuntigamSoleng1997}:
\begin{equation}
\left[i\gamma^{\mu}(x)\partial_{\mu}-i\gamma^{\mu}(x)\Gamma_{\mu}(x)-m\right]\Psi(t,x)=0,\label{9}
\end{equation}
which differs from its flat spacetime counterpart due to the presence
of the additional term $\gamma^{\mu}(x)\Gamma_{\mu}(x)$, accounting
for the geometric effects introduced by the conical defect. The generalized
gamma matrices $\gamma^{\mu}(x)$ satisfy the Clifford algebra $\{\gamma^{\mu},\gamma^{\nu}\}=2g^{\mu\nu}$,
and are expressed in terms of the standard Dirac matrices $\gamma^{a}$
in Minkowski space via the tetrad fields as:
\begin{equation}
\gamma^{\mu}(x)=e_{a}^{\mu}(x)\gamma^{a}.\label{10}
\end{equation}
The tetrads $e_{\mu}^{a}(x)$ fulfill the orthonormality condition:
\begin{equation}
e_{\mu}^{a}(x)e_{\nu}^{b}(x)\eta_{ab}=g_{\mu\nu},\label{11}
\end{equation}
where the indices $\mu,\nu=0,1,2,3$ refer to curved spacetime coordinates,
and $a,b=0,1,2,3$ denote flat spacetime (tetrad) indices. 

The spin connection $\Gamma_{\mu}(x)$ is obtained via:
\begin{equation}
\Gamma_{\mu}(x)=\frac{1}{8}\omega_{\mu ab}(x)[\gamma^{a},\gamma^{b}],\label{12}
\end{equation}
with the spin connection one-forms $\omega_{\mu ab}$ defined as:
\begin{equation}
\omega_{\mu ab}=e_{a\nu}(\partial_{\mu}e_{b}^{\nu}+\Gamma_{\mu\lambda}^{\nu}e_{b}^{\lambda}).\label{13}
\end{equation}
In what follows, we will treat the three dimensional Dirac oscilltor
for each case mentioned above.

\subsection{First case: \foreignlanguage{american}{Dirac Oscillator in a Spinning
Cosmic String Background} with Equal Angular Momentum and Torsion
($j^{2}=0$)}

We consider the Dirac oscillator in the curved background of a spinning
cosmic string with equal temporal and spatial torsion components,
such that $J^{t}=J^{z}=J$, leading to a simplified torsional configuration
$j^{2}=0$. The spacetime geometry is encoded in the tetrad fields:
\begin{equation}
e_{\mu}^{a}(x)=\begin{pmatrix}1 & 0 & 4GJ & 0\\
0 & \cos\phi & -\alpha\rho\sin\phi & 0\\
0 & \sin\phi & \alpha\rho\cos\phi & 0\\
0 & 0 & 4GJ & 1
\end{pmatrix},\quad e_{a}^{\mu}=\begin{pmatrix}1 & 0 & -\frac{4GJ}{\alpha\rho} & 0\\
0 & \cos\phi & \frac{\sin\phi}{\alpha\rho} & 0\\
0 & -\sin\phi & \frac{\cos\phi}{\alpha\rho} & 0\\
0 & 0 & -\frac{4GJ}{\alpha\rho} & 1
\end{pmatrix}.\label{14}
\end{equation}
From these, we obtain the position-dependent gamma matrices in the
curved spacetime:
\begin{equation}
\gamma^{t}=\gamma^{0}-\frac{4GJ}{\alpha\rho}\gamma^{2},\quad\gamma^{\rho}=\cos\phi\,\gamma^{1}+\sin\phi\,\gamma^{2},\quad\gamma^{\phi}=\frac{-\sin\phi\,\gamma^{1}+\cos\phi\,\gamma^{2}}{\alpha\rho},\quad\gamma^{z}=\gamma^{3}-\frac{4GJ}{\alpha\rho}\gamma^{2}.\label{15}
\end{equation}
The Dirac oscillator interaction is introduced via the non-minimal
substitution $\partial_{\rho}\rightarrow\partial_{\rho}+m\omega\gamma^{0}\rho$.
With this, the Dirac equation becomes:
\begin{equation}
\left[\gamma^{0}E-\gamma^{3}k-\gamma^{1}\left(\partial_{\rho}+m\omega\gamma^{0}\rho+\frac{1}{2\rho}\right)-\gamma^{2}\left(\frac{1}{\alpha\rho}\mathcal{J}+\frac{4GJ}{\alpha\rho}(E+k)\right)+m\right]\Psi=0,\label{16}
\end{equation}
where the effective angular momentum operator is 
\begin{equation}
\mathcal{J}=l+\frac{1}{2}-\frac{\alpha}{2}\Sigma^{3}\label{17}
\end{equation}
Using the representation \cite{Carvalho2011}
\begin{equation}
\Psi(t,\rho,\phi,z)=e^{-iEt+i(l+\frac{1}{2}-\frac{\Sigma^{3}}{2})\phi+ikz}\begin{pmatrix}\chi(\rho)\\
\Phi(\rho)
\end{pmatrix}\label{18}
\end{equation}
, we isolate the radial dependence.

We focus on solving the equation for the lower spinor component $\Phi(\rho)$,
by first expressing $\chi$ in terms of $\Phi$:
\begin{equation}
\chi=-\frac{i}{E-m}\left[\sigma^{1}\left(\partial_{\rho}+m\omega\rho+\frac{1}{2\rho}\right)+\mathcal{K}\sigma^{2}+k\sigma^{3}\right]\Phi,\label{19}
\end{equation}
where 
\begin{equation}
\mathcal{K}:=\frac{1}{\alpha\rho}\left(l+\frac{1}{2}-4GJ(E+k)\right)\label{20}
\end{equation}
Substituting this into the lower component equation yields:
\begin{equation}
\left[(E^{2}-m^{2})-\tilde{D}_{+}\tilde{D}_{-}\right]\Phi=0,\label{21}
\end{equation}
with the operators defined as:
\begin{equation}
\tilde{D}_{+}:=\sigma^{1}\left(\partial_{\rho}-m\omega\rho+\frac{1}{2\rho}\right)+\mathcal{K}\sigma^{2}+k\sigma^{3},\quad\tilde{D}_{-}:=\sigma^{1}\left(\partial_{\rho}+m\omega\rho+\frac{1}{2\rho}\right)+\mathcal{K}\sigma^{2}+k\sigma^{3}.\label{22}
\end{equation}
We compute the operator product $\tilde{D}_{+}\tilde{D}_{-}$ explicitly. 

Using standard Pauli matrix identities, the product simplifies to:
\begin{equation}
\tilde{D}_{+}\tilde{D}_{-}=\partial_{\rho}^{2}+\frac{1}{\rho}\partial_{\rho}-m^{2}\omega^{2}\rho^{2}-m\omega+\frac{1}{\rho^{2}}\left(\frac{1}{4}+\frac{1}{\alpha^{2}}\left(l+\frac{1}{2}-4GJ(E+k)\right)^{2}\right)+k^{2}.\label{23}
\end{equation}
The full scalar radial equation for $\Phi(\rho)$ thus becomes:
\begin{equation}
\Phi''+\frac{1}{\rho}\Phi'-m^{2}\omega^{2}\rho^{2}-m\omega+\frac{1}{\rho^{2}}\left(\frac{1}{4}+\frac{1}{\alpha^{2}}\left(l+\frac{1}{2}-4GJ(E+k)\right)^{2}\right)+k^{2}-(E^{2}-m^{2})=0.\label{24}
\end{equation}
Defining the dimensionless variable $x=m\omega\rho^{2}$, the equation
transforms into the canonical form of the confluent hypergeometric
equation \cite{AbramowitzStegun1970,AndrewsAskeyRoy2001}:
\begin{equation}
x\Phi''+\Phi'+\left(-\frac{x}{4}+\frac{\nu^{2}}{4x}-a-\frac{1}{4}\right)\Phi=0,\label{25}
\end{equation}
with
\begin{equation}
a=\frac{E^{2}-m^{2}-k^{2}}{4m\omega},\quad\nu^{2}=\frac{1}{4}+\frac{1}{\alpha^{2}}\left(l+\frac{1}{2}-4GJ(E+k)\right)^{2}.\label{26}
\end{equation}
The solution normal at the origin and vanishing at infinity is:
\begin{equation}
\Phi(x)=x^{\nu/2}e^{-x/2}\,{}_{1}F_{1}(-n,\nu+1,x),\quad n\in\mathbb{N}_{0},\label{27}
\end{equation}
which yields the quantization condition:
\begin{equation}
a=n+\frac{\nu+1}{2}.\label{28}
\end{equation}
Solving for the energy gives the implicit spectrum:
\begin{equation}
E^{2}=m^{2}+k^{2}+4m\omega\left(n+\frac{1}{2}+\frac{1}{2}\sqrt{\frac{1}{4}+\frac{1}{\alpha^{2}}\left(l+\frac{1}{2}-4GJ(E+k)\right)^{2}}\right).\label{29}
\end{equation}
\foreignlanguage{american}{From Eq. (\ref{29}) onward, we impose
the equal-torsion identification $J_{t}=J_{z}\equiv J$. With this
convention, the two torsional couplings merge into a single parameter
that enters the azimuthal sector solely through the energy- and axial-momentum--dependent
shift $\ell+\tfrac{1}{2}-4GJ(E+k)$. Physically, the defect renormalizes
the effective angular index,
\[
\nu^{2}=\tfrac{1}{4}+\frac{1}{\alpha^{2}}\Big[\ell+\tfrac{1}{2}-4GJ(E+k)\Big]^{2},
\]
so the centrifugal barrier and thus the level spacings acquires a
nonlinear dependence on the eigenvalue being determined. Immediate
implications are: (i) degeneracies between distinct $\ell$ are removed;
(ii) for fixed $(\alpha,J,k)$, the variation in $n$ remains monotone,
with spacings distorted relative to flat space; and (iii) the frame-dragging
contribution (time-like torsion) is inseparable from longitudinal
motion, so spectral shifts grow with $|k|$ and with $E$. In the
geometric limit $\alpha\!\to\!1$ and $GJ\!\to\!0$, the spectrum
reduces continuously to the Moshinsky result.}

\begin{figure}
\includegraphics[scale=0.4]{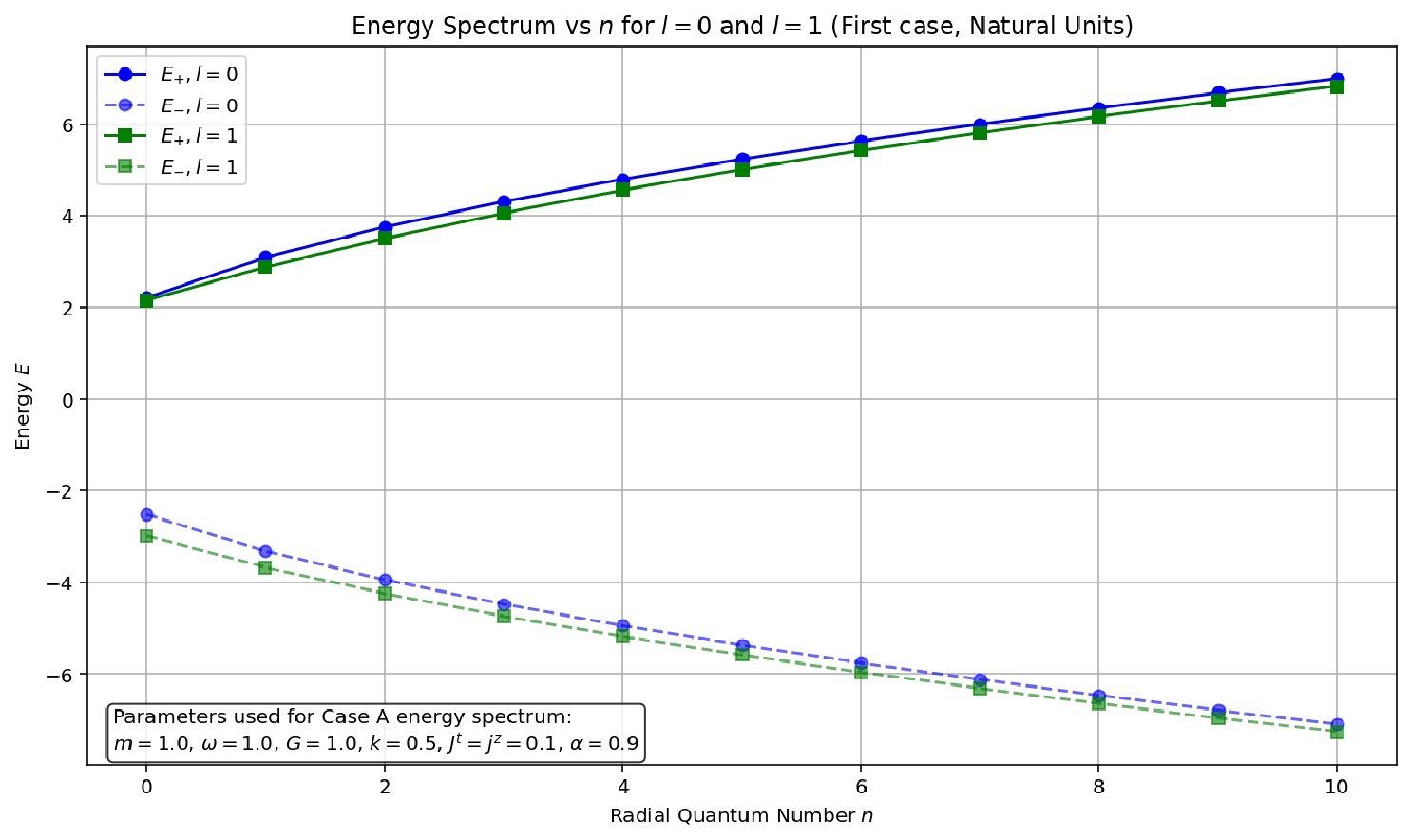}

\caption{\foreignlanguage{american}{Dirac Oscillator in a Spinning Cosmic String Background\foreignlanguage{english}{
with Equal Angular Momentum and Torsion}}}\label{fig:1}
\end{figure}

\selectlanguage{american}%
Figure. \ref{fig:1} displays the energy spectrum $E$ as a function
of the radial quantum number $n$ for angular momentum quantum numbers
$l=0$ and $l=1$, in the case where torsional contributions are balanced---namely,
$J_{t}=J_{z}=J$. The results exhibit a nonlinear growth in energy
with increasing $n$, a hallmark of the Dirac oscillator in geometries
influenced by both curvature and torsion. The upper and lower branches
correspond to positive- and negative-energy (particle and antiparticle)
solutions, respectively. The term $4GJ(E+k)$ introduces an energy-dependent
correction to the effective angular quantum number, breaking the degeneracy
typical of the flat-spacetime limit. Furthermore, the angular deficit
parameter $\alpha$ modifies the curvature of the spectrum, thereby
affecting the centrifugal potential structure.

\selectlanguage{english}%

\subsection{Second case: \foreignlanguage{american}{Dirac Oscillator in a Purely
Spinning Cosmic String Background ($J^{z}=0$)}}

\selectlanguage{american}%
In the second configuration, we consider a purely spinning cosmic
string background, where the temporal component of torsion is retained
while the spatial component vanishes, i.e., $J^{t}\neq0$, $J^{z}=0$.
The tetrad fields reduce to:
\begin{equation}
e_{\mu}^{a}(x)=\begin{pmatrix}1 & 0 & 4GJ^{t} & 0\\
0 & \cos\phi & -\alpha\rho\sin\phi & 0\\
0 & \sin\phi & \alpha\rho\cos\phi & 0\\
0 & 0 & 0 & 1
\end{pmatrix},\quad e_{a}^{\mu}(x)=\begin{pmatrix}1 & 0 & -\frac{4GJ^{t}}{\alpha\rho} & 0\\
0 & \cos\phi & \frac{\sin\phi}{\alpha\rho} & 0\\
0 & -\sin\phi & \frac{\cos\phi}{\alpha\rho} & 0\\
0 & 0 & 0 & 1
\end{pmatrix}.\label{33}
\end{equation}
The corresponding gamma matrices become:
\begin{equation}
\gamma^{t}=\gamma^{0}-\frac{4GJ^{t}}{\alpha\rho}\gamma^{2},\quad\gamma^{\rho}=\cos\phi\gamma^{1}+\sin\phi\gamma^{2},\quad\gamma^{\phi}=\frac{-\sin\phi\gamma^{1}+\cos\phi\gamma^{2}}{\alpha\rho},\quad\gamma^{z}=\gamma^{3}.\label{34}
\end{equation}
Following the same procedure of decoupling the Dirac equation using
the lower spinor component $\Phi(\rho)$, and isolating the second-order
differential equation, we find:
\begin{equation}
\Phi''+\frac{1}{\rho}\Phi'-m^{2}\omega^{2}\rho^{2}-m\omega+\frac{1}{\rho^{2}}\left(\frac{1}{4}+\frac{1}{\alpha^{2}}\left(l+\frac{1}{2}-4GJ^{t}(E+k)\right)^{2}\right)+k^{2}-(E^{2}-m^{2})=0.\label{35}
\end{equation}
After the change of variable $x=m\omega\rho^{2}$, this transforms
into the confluent hypergeometric form:
\begin{equation}
x\Phi''+\Phi'+\left(-\frac{x}{4}+\frac{\nu^{2}}{4x}-a-\frac{1}{4}\right)\Phi=0,\label{36}
\end{equation}
with parameters:
\begin{equation}
a=\frac{E^{2}-m^{2}-k^{2}}{4m\omega},\quad\nu^{2}=\frac{1}{4}+\frac{1}{\alpha^{2}}\left(l+\frac{1}{2}-4GJ^{t}(E+k)\right)^{2}.\label{37}
\end{equation}
The normalized solution 
\begin{equation}
\Phi(x)=x^{\nu/2}e^{-x/2}{}_{1}F_{1}(-n,\nu+1,x)\label{38}
\end{equation}
 leads to the quantization condition:
\begin{equation}
E^{2}=m^{2}+k^{2}+4m\omega\left(n+\frac{1}{2}+\frac{1}{2}\sqrt{\frac{1}{4}+\frac{1}{\alpha^{2}}\left(l+\frac{1}{2}-4GJ^{t}(E+k)\right)^{2}}\right).\label{39}
\end{equation}
This case retains the structure of case A but removes the spatial
torsion term, isolating the influence of time-like angular momentum.
This isolates a pure frame-dragging effect: the deformation of the
angular channel depends on $E+k$ (implicit eigenvalue problem) with
no spatial-torsion (screw) contribution. Qualitatively, increasing
$|J_{t}|$ strengthens the $\ell$-dependent splitting at small $|\ell|$
and pushes the ground state upward; at large $|\ell|$ the relative
effect weakens as the $\ell^{2}/\alpha^{2}$ piece dominates. Again,
flat space is recovered for $GJ_{t}\!\to\!0$ and $\alpha\!\to\!1$.

\begin{figure}
\selectlanguage{english}%
\includegraphics[scale=0.4]{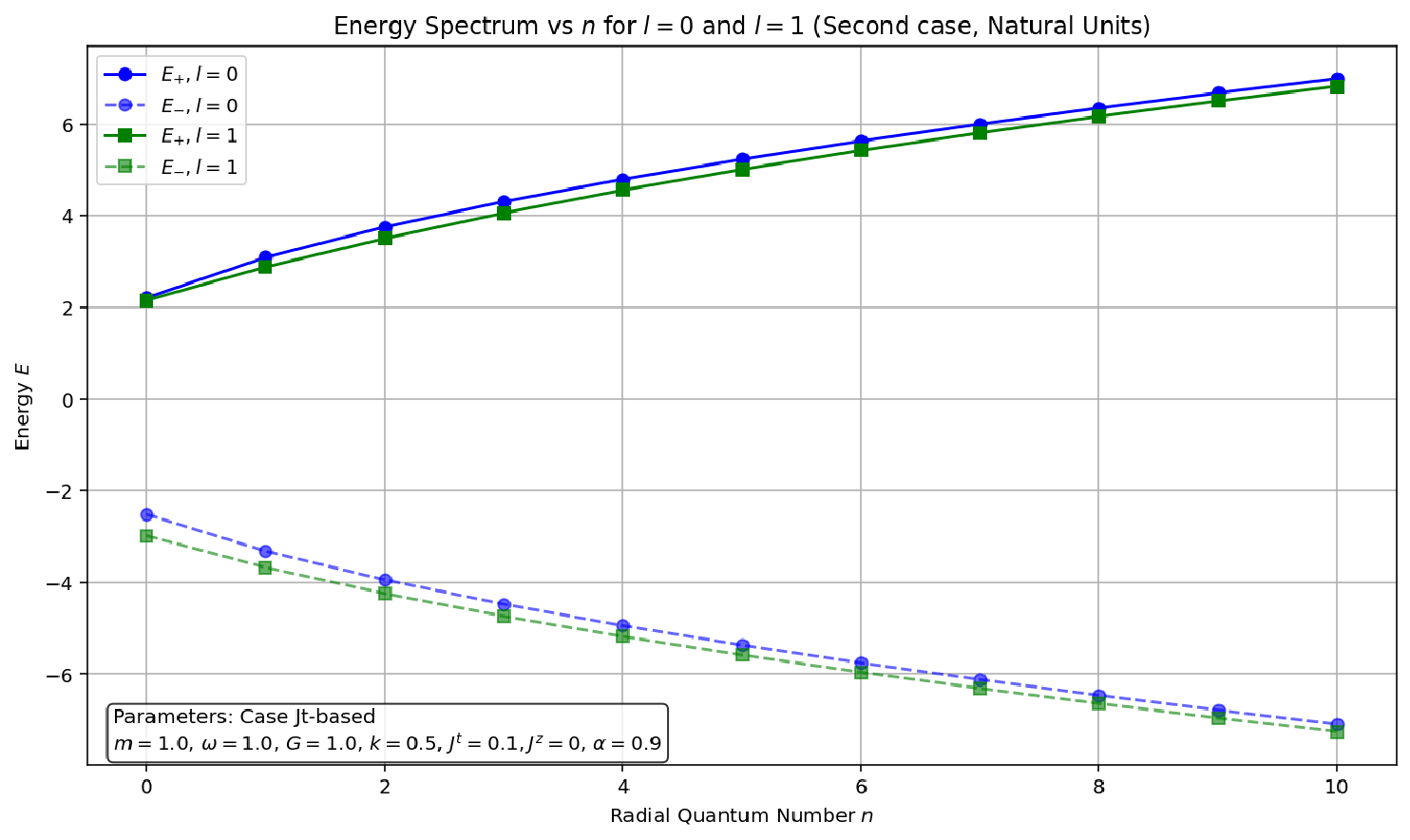}

\caption{\foreignlanguage{american}{Dirac Oscillator in a Purely Spinning Cosmic String Background}}\label{fig:2}
\selectlanguage{american}%
\end{figure}

Figure. \ref{fig:2} illustrates the energy spectrum for a configuration
characterized by purely temporal torsion, where $J_{z}=0$. The energy
levels maintain a nonlinear dependence on $n$, governed entirely
by the time-like torsional component $J^{t}$. The absence of spatial
torsion simplifies the angular deformation, isolating the influence
of frame-dragging on the relativistic dynamics. The comparison across
different values of $l$ demonstrates how the torsional parameter
$J^{t}$ modifies the angular sector while preserving the energy-dependent
nature of the quantization condition.
\selectlanguage{english}%

\subsection{Third case\foreignlanguage{american}{: Dirac Oscillator in a Cosmic
String Background with Screw Dislocations ($J^{t}=0$)}}

\selectlanguage{american}%
In the third case, the background spacetime consists of a cosmic string
with screw dislocation, where $J^{t}=0$ and $J^{z}\neq0$. This reflects
purely spatial torsion along the $z$-axis. The tetrads simplify accordingly:
\begin{equation}
e_{\mu}^{a}(x)=\begin{pmatrix}1 & 0 & 0 & 0\\
0 & \cos\phi & -\alpha\rho\sin\phi & 0\\
0 & \sin\phi & \alpha\rho\cos\phi & 0\\
0 & 0 & 4GJ^{z} & 1
\end{pmatrix},\quad e_{a}^{\mu}(x)=\begin{pmatrix}1 & 0 & 0 & 0\\
0 & \cos\phi & \frac{\sin\phi}{\alpha\rho} & 0\\
0 & -\sin\phi & \frac{\cos\phi}{\alpha\rho} & 0\\
0 & 0 & -\frac{4GJ^{z}}{\alpha\rho} & 1
\end{pmatrix}.\label{40}
\end{equation}
The corresponding gamma matrices read:
\begin{equation}
\gamma^{t}=\gamma^{0},\quad\gamma^{\rho}=\cos\phi\gamma^{1}+\sin\phi\gamma^{2},\quad\gamma^{\phi}=\frac{-\sin\phi\gamma^{1}+\cos\phi\gamma^{2}}{\alpha\rho},\quad\gamma^{z}=\gamma^{3}-\frac{4GJ^{z}}{\alpha\rho}\gamma^{2}.\label{41}
\end{equation}
As in the previous cases, solving the Dirac equation using the lower
component $\Phi(\rho)$ leads to:
\begin{equation}
\Phi''+\frac{1}{\rho}\Phi'-m^{2}\omega^{2}\rho^{2}-m\omega+\frac{1}{\rho^{2}}\left(\frac{1}{4}+\frac{1}{\alpha^{2}}\left(l+\frac{1}{2}-4GJ^{z}k\right)^{2}\right)+k^{2}-(E^{2}-m^{2})=0.\label{42}
\end{equation}
Upon substitution $x=m\omega\rho^{2}$, this becomes:
\begin{equation}
x\Phi''+\Phi'+\left(-\frac{x}{4}+\frac{\nu^{2}}{4x}-a-\frac{1}{4}\right)\Phi=0,\label{43}
\end{equation}
where now:
\begin{equation}
a=\frac{E^{2}-m^{2}-k^{2}}{4m\omega},\quad\nu^{2}=\frac{1}{4}+\frac{1}{\alpha^{2}}\left(l+\frac{1}{2}-4GJ^{z}k\right)^{2}.\label{44}
\end{equation}
The solution and quantization condition are formally identical in
structure to the previous cases, yielding:
\begin{equation}
E^{2}=m^{2}+k^{2}+4m\omega\left(n+\frac{1}{2}+\frac{1}{2}\sqrt{\frac{1}{4}+\frac{1}{\alpha^{2}}\left(l+\frac{1}{2}-4GJ^{z}k\right)^{2}}\right).\label{45}
\end{equation}
Unlike Cases A and B, this configuration generates torsional correction
terms independent of energy $E$, depending instead only on the longitudinal
momentum $k$. This result emphasizes the physical distinction between
time-like and space-like torsion: while the former induces an implicit
nonlinear eigenvalue problem, the latter introduces explicit momentum-dependent
deformation of the angular eigenvalues. In addition, because $E$
no longer appears under the square-root shift, the spectrum is explicit
in $E$. The screw dislocation acts like a $k$-controlled Aharonov--Bohm--type
twist of the angular index, splitting levels with different $\ell$
in a way that grows with $|k|$ but does not feed back on $E$. This
cleanly separates the roles of longitudinal transport (through $k$)
and angular dynamics (through $\ell$ and $\alpha$).

\begin{figure}
\selectlanguage{english}%
\includegraphics[scale=0.4]{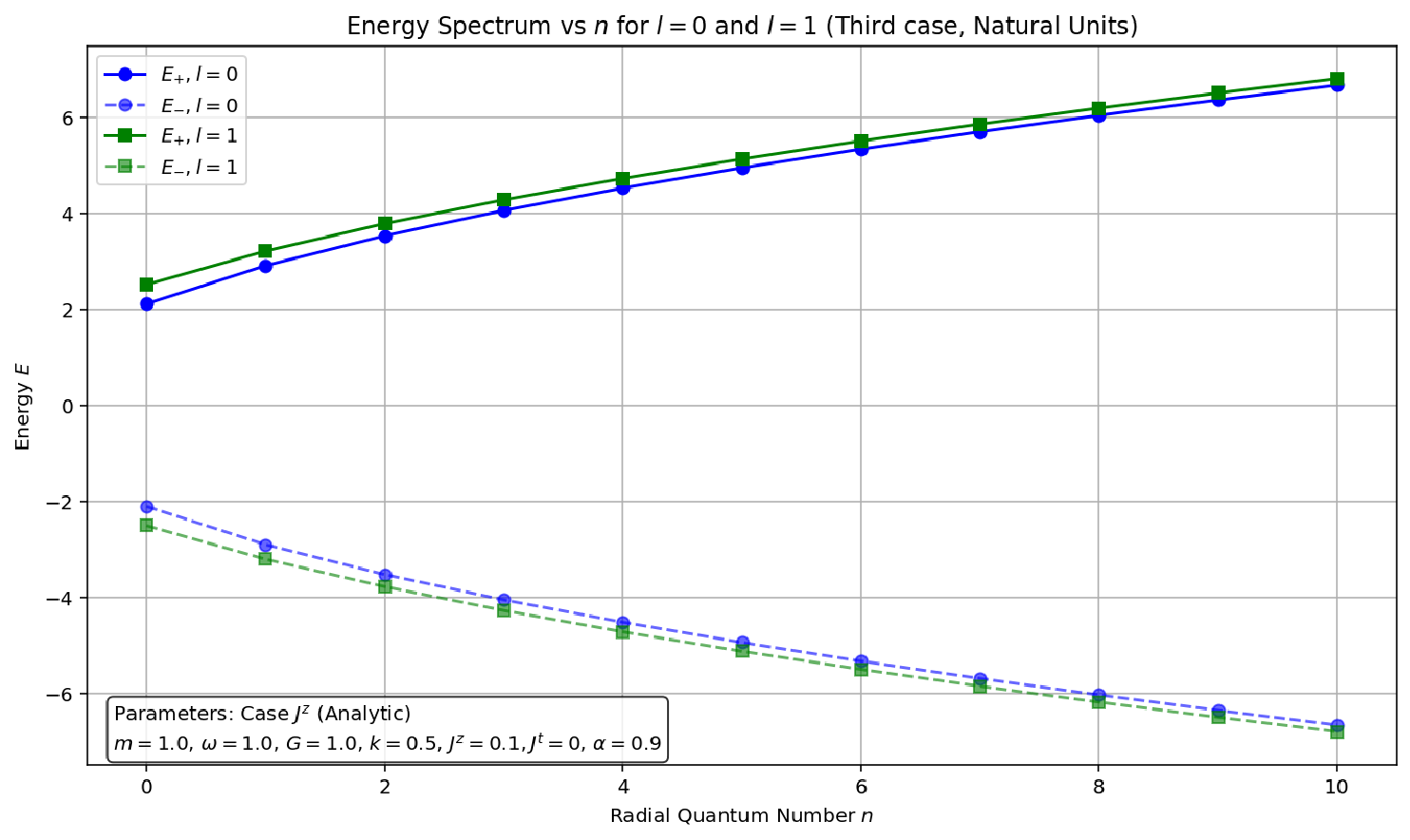}

\caption{\foreignlanguage{american}{Dirac Oscillator in a Cosmic String Background with Screw Dislocations}}\label{fig:3}
\selectlanguage{american}%
\end{figure}

Figure. \ref{fig:3} presents the energy spectrum for the case involving
purely spatial torsion ($J^{t}=0$, $J^{z}\neq0$), corresponding
to a screw dislocation along the $z$-axis. In this setting, the correction
to the angular quantum number arises from the term $4GJ^{z}k$, establishing
a direct dependence on the longitudinal momentum $k$. Unlike the
previous cases, the energy levels are expressed explicitly and do
not involve recursive dependence on $E$. This distinction underscores
the qualitative difference between space-like and time-like torsion:
spatial torsion introduces momentum-dependent but energy-independent
modifications to the angular structure. As $n$ increases, the energy
separation between different $l$ states becomes more pronounced,
revealing the geometric influence on orbital dynamics.

\subsection{General case\foreignlanguage{english}{: Dirac Oscillator in a Torsion
and Curvature-Modified Spacetime}}

\selectlanguage{english}%
We consider the dynamics of the Dirac oscillator in a nontrivial spacetime
background incorporating both curvature and torsion. The metric is
specified in matrix form as
\begin{equation}
g_{\mu\nu}=\begin{pmatrix}-1 & 0 & -4GJ^{t} & 0\\
0 & 1 & 0 & 0\\
-4GJ^{t} & 0 & \alpha^{2}\rho^{2}-16G^{2}\left[(J^{t})^{2}-(J^{z})^{2}\right] & 4GJ^{z}\\
0 & 0 & 4GJ^{z} & 1
\end{pmatrix},\label{46}
\end{equation}
This metric includes both off-diagonal and curvature-corrected diagonal
components, and is consistent with cosmic string-like sources with
intrinsic spin and dislocation.

To construct a spinor formalism in this background, we seek a tetrad
field $e_{\mu}^{a}(x)$ satisfying $g_{\mu\nu}=e_{\mu}^{a}e_{\nu}^{b}\eta_{ab}$,
where $\eta_{ab}=\mathrm{diag}(-1,1,1,1)$ is the Minkowski metric
in the local Lorentz frame. 

A compatible choice is
\begin{equation}
e_{\mu}^{a}(x)=\begin{pmatrix}1 & 0 & 4GJ^{t} & 0\\
0 & 1 & 0 & 0\\
0 & 0 & \alpha\rho & 0\\
0 & 0 & 4GJ^{z} & 1
\end{pmatrix},\quad e_{a}^{\mu}(x)=\begin{pmatrix}1 & 0 & -\dfrac{4GJ^{t}}{\alpha\rho} & 0\\
0 & 1 & 0 & 0\\
0 & 0 & \dfrac{1}{\alpha\rho} & 0\\
0 & 0 & -\dfrac{4GJ^{z}}{\alpha\rho} & 1
\end{pmatrix}.\label{47}
\end{equation}
From this tetrad, the curved-space Dirac matrices are constructed
via $\gamma^{\mu}(x)=e_{a}^{\mu}(x)\gamma^{a}$, yielding:
\begin{equation}
\gamma^{t}=\gamma^{0}-\frac{4GJ^{t}}{\alpha\rho}\gamma^{2},\quad\gamma^{\rho}=\gamma^{1},\quad\gamma^{\phi}=\frac{1}{\alpha\rho}\gamma^{2},\quad\gamma^{z}=\gamma^{3}-\frac{4GJ^{z}}{\alpha\rho}\gamma^{2}.\label{48}
\end{equation}
The Dirac oscillator is introduced by the standard substitution in
the radial derivative,
\begin{equation}
\partial_{\rho}\rightarrow\partial_{\rho}+m\omega\beta\rho,\quad\beta=\gamma^{0},\label{49}
\end{equation}
into the covariant Dirac equation. Using the spinor ansatz
\begin{equation}
\Psi(t,\rho,\phi,z)=e^{-iEt+i(l+\frac{1}{2}-\frac{\Sigma^{3}}{2})\phi+ikz}\begin{pmatrix}\chi(\rho)\\
\Phi(\rho)
\end{pmatrix},\label{50}
\end{equation}
the radial part of the Dirac equation becomes, after eliminating the
upper component $\chi(\rho)$,
\begin{equation}
\left[E^{2}-m^{2}-\tilde{D}_{+}\tilde{D}_{-}\right]\Phi=0,\label{51}
\end{equation}
where $\tilde{D}_{\pm}$ are the radial operators,
\begin{equation}
\tilde{D}_{\pm}=\sigma^{1}\left(\partial_{\rho}\pm m\omega\rho+\frac{1}{2\rho}\right)+\mathcal{K}\sigma^{2}+k\sigma^{3},\label{52}
\end{equation}
and the effective angular coupling is
\begin{equation}
\mathcal{K}=\frac{1}{\alpha\rho}\left(l+\frac{1}{2}-4GJ^{t}(E+k)-4GJ^{z}k\right).\label{53}
\end{equation}
Explicitly evaluating $\tilde{D}_{+}\tilde{D}_{-}$, we obtain the
second-order scalar differential equation for the lower spinor component
$\Phi(\rho)$:
\begin{equation}
\Phi''+\frac{1}{\rho}\Phi'-m^{2}\omega^{2}\rho^{2}-m\omega+\frac{1}{\rho^{2}}\left[\frac{1}{4}+\frac{1}{\alpha^{2}}\left(l+\frac{1}{2}-4GJ^{t}(E+k)-4GJ^{z}k\right)^{2}\right]+k^{2}-(E^{2}-m^{2})=0.\label{54}
\end{equation}
Introducing the dimensionless variable $x=m\omega\rho^{2}$, this
equation transforms into a confluent hypergeometric-type equation:
\begin{equation}
x\Phi''+\Phi'+\left(-\frac{x}{4}+\frac{\nu^{2}}{4x}-a-\frac{1}{4}\right)\Phi=0,\label{55}
\end{equation}
with parameters:
\begin{equation}
a=\frac{E^{2}-m^{2}-k^{2}}{4m\omega},\quad\nu^{2}=\frac{1}{4}+\frac{1}{\alpha^{2}}\left(l+\frac{1}{2}-4GJ^{t}(E+k)-4GJ^{z}k\right)^{2}.\label{56}
\end{equation}
The normalized solution is given by:
\begin{equation}
\Phi(x)=x^{\nu/2}e^{-x/2}{}_{1}F_{1}(-n,\nu+1,x),\quad n=0,1,2,\dots,\label{57}
\end{equation}
with the quantization condition obtained from the polynomial truncation:
\begin{equation}
a=n+\frac{\nu+1}{2}.\label{58}
\end{equation}
This leads to the final energy spectrum:
\begin{equation}
E^{2}=m^{2}+k^{2}+4m\omega\left(n+\frac{1}{2}+\frac{1}{2}\sqrt{\frac{1}{4}+\frac{1}{\alpha^{2}}\left(l+\frac{1}{2}-4GJ^{t}(E+k)-4GJ^{z}k\right)^{2}}\right).\label{59}
\end{equation}
This expression reflects the total influence of both curvature (disclination)
and torsion (dislocation and spinning effects), leading to a highly
nontrivial energy spectrum that is implicit in the energy $E$ and
exhibits strong coupling between geometry, spin, and momentum. \foreignlanguage{american}{The
effective angular index now carries both a time-like part that is
implicit in $E$ and a space-like part controlled by $k$. Time-like
torsion (via $J_{t}$) produces the nonlinear, self-consistent shift
typical of Cases A/B; space-like torsion (via $J_{z}$) adds a linear
$k$-dependent bias as in Case C. Together with the conical deficit
($\alpha<1$), these ingredients (i) remove flat-space degeneracies,
(ii) skew level spacings across $\ell$, and (iii) couple longitudinal
motion to the angular barrier. All special limits (turning off $J_{t}$,
$J_{z}$, or restoring $\alpha\!\to\!1$) reproduce the appropriate
sub-cases and finally the Moshinsky result.}

\begin{figure}
\includegraphics[scale=0.4]{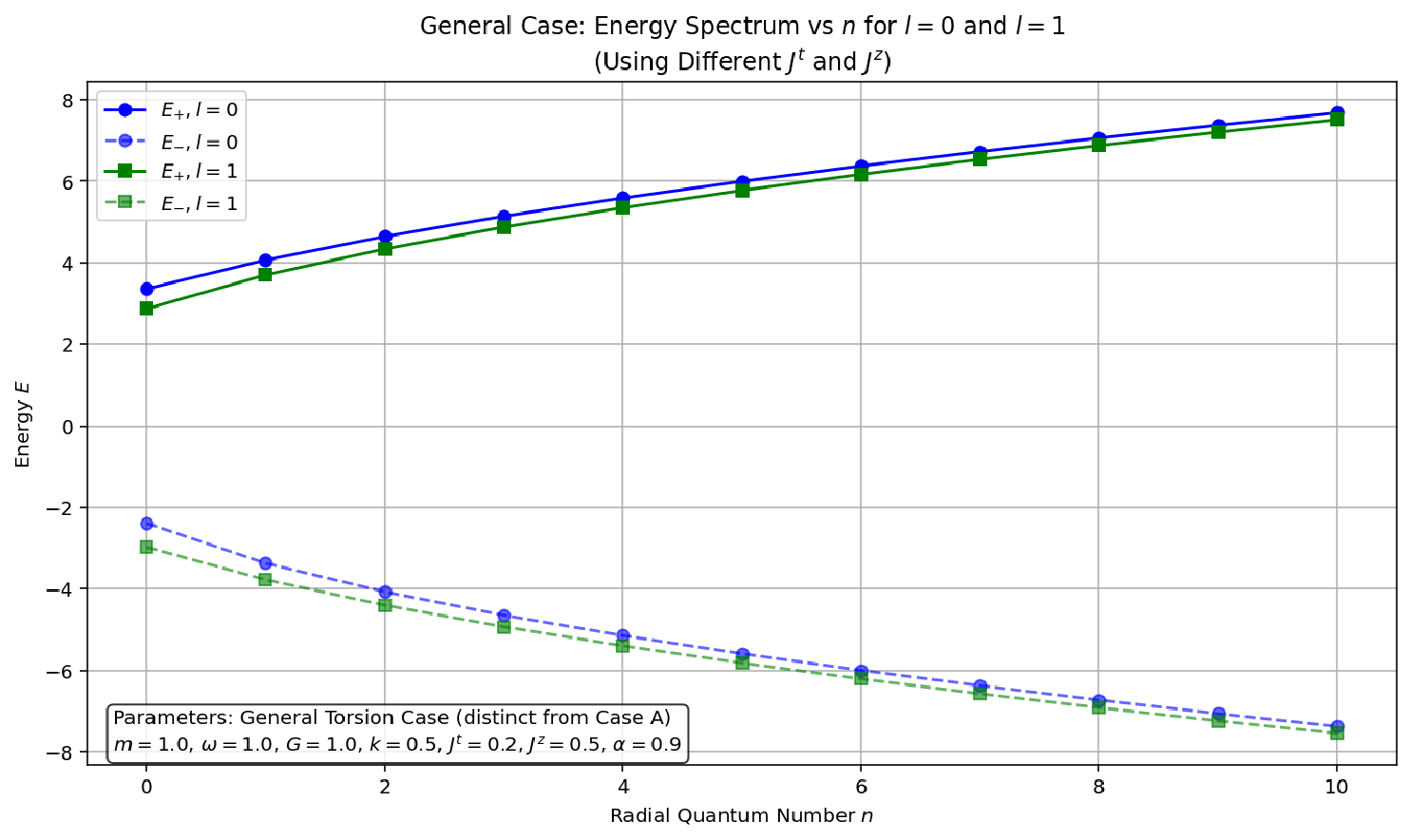}

\caption{Dirac Oscillator in a Torsion and Curvature-Modified Spacetime}\label{fig:4}
\end{figure}

\selectlanguage{american}%
Figure. \ref{fig:4} synthesizes the effects of both curvature (encoded
by the deficit parameter $\alpha$) and torsion (via $J^{t}$ and
$J^{z}$) on the Dirac oscillator spectrum. The plot shows the energy
variation with respect to $n$ for $l=0$ and $l=1$, under parameter
choices distinct from previous cases to highlight the joint impact.
The energy spectrum is implicitly defined, with a complex dependence
on both the energy $E$ and longitudinal momentum $k$, reflecting
the nontrivial interplay among curvature, torsion, and spinor dynamics.
The resulting spectral structure illustrates how the combination of
spinning and dislocated spacetime defects induces cumulative shifts
in the energy levels, leading to pronounced degeneracy lifting and
intricate spectral behavior as $n$ increases.

Finally, since $g_{\phi\phi}=\alpha^{2}\rho^{2}-16G^{2}\!\left[(J_{t})^{2}-(J_{z})^{2}\right]$,
the metric requires $g_{\phi\phi}>0$. When $|J_{t}|>|J_{z}|$, this
condition implies a minimal admissible radius
\begin{equation}
\rho_{c}=\frac{4G}{\alpha}\sqrt{(J_{t})^{2}-(J_{z})^{2}},\label{59-1}
\end{equation}
and the radial problem must be posed with boundary conditions at $\rho=\rho_{c}$.
This effective “hard wall” strengthens the centrifugal barrier most
noticeably at small $|\ell|$ and drives the spectrum upward. By contrast,
for $J_{t}=0$ or $|J_{t}|\le|J_{z}|$ the full domain $(0,\infty)$
is available. These geometric constraints are compatible with solvability
of the radial equation in terms of confluent hypergeometric functions
and with the quantization prescription employed for all four spectra
above.

In the combined limit $GJ_{t},\,GJ_{z}\!\to\!0$ and $\alpha\!\to\!1$,
every spectrum continuously reduces to
\begin{equation}
E^{2}=m^{2}+k^{2}+4m\omega\!\left(n+\tfrac{1}{2}+\tfrac{1}{2}\lvert\ell+\tfrac{1}{2}\rvert\right),\label{59-2}
\end{equation}
thereby recovering the standard two--dimensional Dirac-oscillator
result and its familiar degeneracies. For weak defects $\lvert1-\alpha\rvert\ll1$
and small $GJ_{t},GJ_{z}$ the leading spectral shifts are governed
by linear variations of the effective angular index, offering a clear,
testable perturbative signature that distinguishes curvature (via
$\alpha$) from torsion (via $J_{t},J_{z}$).

\section{Results and discussion}

\selectlanguage{english}%
The analytical solutions derived for the Dirac oscillator in the presence
of a spinning cosmic string with curvature and torsion exhibit a unified
spectral structure governed by confluent hypergeometric functions.
Despite the incorporation of nontrivial topological features, such
as angular deficit and screw dislocation, the system remains exactly
solvable. In all geometrical configurations examined, the energy spectrum
reflects the influence of the background geometry through energy-
and momentum-dependent modifications to the angular momentum.

In the first configuration, characterized by balanced temporal and
spatial torsion contributions, the energy levels are implicitly defined
and incorporate both energy and momentum within a modified angular
quantum number. The inclusion of the term proportional to $4GJ(E+k)$
in the angular contribution induces a nonlinear coupling between the
spinor field and the torsional geometry. This structure deviates from
the linear form typical of flat-spacetime systems and highlights a
novel interaction between relativistic spin and spacetime torsion.
Consequently, the quantization condition becomes self-referential
in $E$, illustrating that the effective angular momentum is no longer
a fixed quantum number but a dynamic quantity dependent on the particle's
motion in a torsion-enriched background.

The second configuration, featuring only the temporal torsion component,
simplifies the spectral structure while preserving the essential energy
dependence. The elimination of spatial torsion yields a clearer distinction
between angular and translational contributions, although the torsional
term $4GJ^{t}E$ persists in modifying the angular quantum number.
This scenario can be interpreted as a pure frame-dragging effect,
demonstrating how temporal torsion alone can alter the oscillator's
spectrum without introducing explicit momentum dependence in the angular
sector. The eigenvalues remain implicit but afford a more straightforward
physical interpretation of the manner in which rotational defects
impact spin dynamics.

In contrast, the third configuration examines a background dominated
by spatial torsion, corresponding to a screw dislocation with $J^{t}=0$.
Here, the spectrum is explicit in $E$ and incorporates a deformation
dependent solely on the longitudinal momentum $k$. This reflects
a decoupling between energy and the torsional correction, with the
angular quantum number modified by the spatial geometry via the term
$4GJ^{z}k$. Such a finding differentiates space-like torsion from
time-like torsion: the former influences the particle's orbital behavior
in a momentum-dependent yet energy-independent manner, whereas the
latter engenders a nonlinear interplay between spin and energy. Although
structurally akin to the preceding cases, the spectral expression
thus arises from a fundamentally distinct physical mechanism underlying
the angular deformation.

The general case, encompassing both curvature and torsion, yields
an implicit spectral equation that integrates the aforementioned configurations.
The aggregate torsional contribution manifests in the angular shift
as a amalgamation of $J^{t}$ and $J^{z}$, resulting in a deformation
proportional to $4GJ^{t}E+4GJ^{z}k$. Accordingly, the angular quantum
number is supplanted by a dynamic, geometry-dependent function that
concurrently encapsulates the effects of spacetime rotation, dislocation,
and conical topology. The energy spectrum in this instance manifests
a robust coupling between the oscillator's intrinsic dynamics and
the ambient geometry, thereby extending the Dirac oscillator to curved
and torsion-laden spacetimes.

A meaningful comparison arises with the flat-space Dirac oscillator
introduced by Moshinsky and Szczepaniak, who proposed a linear modification
to the Dirac equation through the substitution $\vec{p}\rightarrow\vec{p}-im\omega\beta\vec{r}$.
In their formulation, the squared Dirac equation yields an effective
Hamiltonian consisting of a harmonic oscillator term plus a strong
spin-orbit coupling. In the nonrelativistic limit, the energy levels
become:
\begin{equation}
\mathcal{E}=\hbar\omega\left(2n+l+\frac{3}{2}\right)-2\omega\vec{L}\cdot\vec{S},\label{30-1}
\end{equation}
and the relativistic spectrum reads:
\begin{equation}
E^{2}=m^{2}c^{4}+\hbar^{2}\omega^{2}\left[2N+1\pm2j\right].\label{31-1}
\end{equation}
Our curved-spacetime spectrum generalizes this result. In the limit
$\alpha\to1$, $GJ\to0$, and $k\to0$, the energy reduces to:
\begin{equation}
E^{2}=m^{2}+4m\omega\left(n+\frac{1}{2}+\frac{1}{2}|l+\tfrac{1}{2}|\right),\label{32-1}
\end{equation}
which is precisely the flat-space 2D Dirac oscillator spectrum. The
replacement of the spin-orbit interaction $2\omega\vec{L}\cdot\vec{S}$
by a geometry- and energy-dependent deformation of angular momentum
reflects a fundamental extension of the original Moshinsky oscillator:
the curvature and torsion of spacetime now dynamically alter the spinor
structure of the relativistic wavefunction. This model not only recovers
Moshinsky’s result as a special case but reveals new relativistic
regimes where geometry and quantum fields are intrinsically coupled.

These findings can be directly contrasted with the standard Dirac
oscillator spectrum in flat spacetime, as formulated by Moshinsky
and Szczepaniak. In their framework, the energy levels are given by
\begin{equation}
E^{2}=m^{2}+k^{2}+4m\omega\left(n+\frac{1}{2}+\frac{1}{2}|\kappa|\right),\label{60}
\end{equation}
where $\kappa=\pm(j+1/2)$ encapsulates the spin-orbit coupling, and
the angular quantum number is static and integer-valued. This spectrum,
derived within Minkowski geometry, exhibits well-established degeneracies
and an algebraic structure linked to the $\mathfrak{so}(4)$ or $\mathfrak{so}(3,1)$
symmetry. In the generalized model presented herein, however, the
term under the square root is supplanted by a dynamically deformed
angular quantity
\begin{equation}
\sqrt{\frac{1}{4}+\frac{1}{\alpha^{2}}\left(l+\frac{1}{2}-4GJ^{t}E-4GJ^{z}k\right)^{2}},\label{61}
\end{equation}
which functions as an effective, energy- and momentum-dependent extension
of $|\kappa|$. The degeneracy inherent in Moshinsky’s formulation
is thereby disrupted. The previously fixed labels $j=\ell\pm\frac{1}{2}$
are replaced by a continuous angular deformation modulated by spacetime
geometry. This alteration not only adjusts the energy level spacing
but also induces torsion- and curvature-dependent splittings among
levels sharing identical quantum numbers $(n,l)$, thus embedding
geometric information within the spectral profile.

Notably, in the limit where all geometric deformations are nullified---i.e.,
$\ensuremath{J^{t}\to0},\ensuremath{J^{z}\to0}$, $\alpha\to1$, and
$k\to0$---the energy spectrum reverts precisely to the Moshinsky
result. This affirms the robustness of the flat-spacetime oscillator
as a special case of a broader curved-spacetime framework. Nevertheless,
the integration of curvature and torsion engenders corrections with
profound physical ramifications: the quantized energy levels of relativistic
particles become attuned to global spacetime topology and torsional
anomalies.

Therefore, the model delineated in this study not only broadens the
applicability of the Dirac oscillator to curved spacetimes but also
furnishes a paradigm whereby spectral perturbations can elucidate
underlying geometric and topological attributes. These insights hold
promise for applications in analog condensed-matter systems, such
as graphene or optical lattices, where effective geometries emulate
curvature and torsion, potentially enabling experimental corroboration
of these theoretical prognoses. In this expansive purview, the Dirac
oscillator transcends mere mathematical abstraction, emerging as a
diagnostic instrument for interrogating the geometry of quantum systems.

\section{Potential Experimental Implementations}

The Dirac oscillator (DO) is obtained via the nonminimal substitution
\begin{equation}
\mathbf{p}\;\longrightarrow\;\mathbf{p}-i\,m\omega\,\beta\,\mathbf{r},\label{60-1}
\end{equation}
which preserves linearity in momentum and yields exactly (or quasi-exactly)
solvable structures with a transparent algebra. Although extensively
explored theoretically, experimental demonstrations of Dirac-oscillator
dynamics have emerged only recently, notably in arrays of microwave
resonators\cite{Franco-Villafane2013}. Related Dirac dynamics have
also been realized with trapped ions and in Dirac materials, including
graphene.

Graphene provides a controlled realization of $(2+1)$-dimensional
Dirac quasiparticles: near the $K/K'$ valleys, low-energy excitations
obey an effective massless Dirac equation on the honeycomb lattice,
accounting for the material’s distinctive transport and mechanical
responses. This setting enables field-theoretic concepts to be explored
at experimentally accessible energy scales.

Topological defects, particularly disclinations generated by removing
or inserting angular sectors, introduce localized curvature (and,
in elastic continua, effective torsion). In graphene, these defects
correspond to non-hexagonal rings (e.g., pentagons/heptagons) and
can reshape a flat sheet into cones, fullerene-type structures, wormholes,
and related geometries. The geometric theory of defects due to Katanaev
and Volovich frames these features by encoding defect content into
curvature/torsion fields of an elastic manifold, allowing low-energy
carriers to be modeled as Dirac spinors propagating on a curved background
with emergent gauge connections. Within this continuum description,
disclinations act as conical singularities with deficit angle $\delta=2\pi(1-\alpha)$,
where $0<\alpha\le1$. They modify interference and transport via
Berry/Aharonov--Bohm--type phases for transported spinors and shift
spectral features such as Landau levels in external magnetic fields.
Beyond single-particle spectra, the associated holonomies have been
proposed as resources for geometric control in graphene-based quantum
devices, and related geometric-phase effects have been analyzed using
Kaluza--Klein--inspired elastic geometries (for more details see
Refs. \cite{BakkeFurtado2009,BakkeFurtado2010,Garcia2025EPJPlus,Carvalho2011,Bueno2025AOP,MarquesEtAl2005,Oliveira2021AOP}).
In summary, graphene realizes a versatile Dirac medium in which the
spectral and phase consequences of topological defects can be quantified,
directly linking defect geometry to measurable electronic observables.

In $(2+1)$ dimensions, Bermúdez \textbackslash emph\{et al.\} \cite{Bermudez2007,Bermudez2008}
derived exact solutions and established an explicit mapping to Jaynes--Cummings/anti--Jaynes--Cummings
(JC/AJC) models, thereby importing quantum-optical intuition and opening
experimental routes. Dirac-type dynamics have since been emulated
in trapped ions and microwave-resonator arrays, with state preparation
and readout sufficient to probe relativistic effects \cite{Franco-Villafane2013,Yang2020,Canas2025,Zhang2025,Gerritsma2010Nature,Lamata2011,Blatt2012}.

Embedding the DO in a cosmic-string--type background introduces a
conical geometry characterized by the same deficit parameter $\alpha$
that models a disclination in graphene. The corresponding zweibein
and spin connection shift the effective angular-momentum index and
imprint an Aharonov--Bohm-like phase, thereby deforming the DO spectrum
and eigenstates. In disclinated graphene, the elastic geometry produces
an equivalent conical metric together with emergent gauge fields (including
strain-induced pseudomagnetic fields) acting on the valley--spinor
structure. Consequently, the {*}{*}DO on a conical spacetime and the
DO in disclinated graphene share the following structural elements:
\begin{itemize}
\item The JC/AJC-compatible ladder structure inherited from the oscillator
coupling;
\item An $\alpha$-dependent modification of the effective angular momentum
and boundary conditions;
\item Topological phases: defect-generated holonomies that affect quantization
and selection rules.
\end{itemize}
These ingredients yield (i) defect-dependent lifting of degeneracies,
(ii) shifts of Landau-like levels in external or pseudo-magnetic fields,
and (iii) characteristic modifications of radial probability profiles
and transition amplitudes. Experimentally, graphene nanocones or patterned
disclination arrays emulate the conical metric, while strain engineering
tunes pseudogauge fields and the effective $\alpha$. In parallel,
trapped-ion simulators implement the same algebra via Jaynes--Cummings
(JC)/anti--Jaynes--Cummings(AJC) \cite{JaynesCummings1963,Knight2024JOSAB,ShoreKnight1993}
(couplings with synthetic gauge fields and programmable boundary conditions,
offering a complementary high-fidelity platform to interrogate DO-with-defects
physics.

Overall, graphene with disclinations provides a condensed-matter analogue
of the Dirac oscillator in a cosmic-string background. The JC/AJC
mapping unifies the theoretical description across graphene, trapped-ion,
and photonic platforms, clarifying how topological defects are transduced
into measurable relativistic-like spectral signatures without requiring
ultra-high energy scales.

This perspective is further supported by the recent work of Majumdar
et al. \cite{Majumdar2025}. Their study confirms the presence of
a Dirac fluid---a distinctive quantum state---in high-purity graphene
samples. The research demonstrates a pronounced deviation from the
Wiedemann-Franz law, with electrical and thermal conductivities displaying
an inverse correlation, exceeding the expected ratio by over 200 times
at low temperatures. This decoupling is governed by the quantum of
conductance, a universal quantum constant, observed at the Dirac point,
where graphene exhibits a transition between metallic and insulating
behavior. At this critical point, electrons display collective, fluid-like
dynamics with minimal viscosity, akin to a quark-gluon plasma. These
findings underscore graphene’s utility as a cost-effective platform
for investigating high-energy physics phenomena, such as black-hole
thermodynamics, and its potential for developing quantum sensors capable
of detecting minute electrical signals and magnetic fields.

\section{Conclusion}

We have presented an exact analytical solution of the covariant Dirac-oscillator
in a spinning cosmic-string spacetime that simultaneously incorporates
conical curvature (disclination) and torsion (spinning and screw dislocation).
Working in a tetrad frame, the radial problem maps to the confluent-hypergeometric
equation, yielding normalizable states and a closed quantization rule.
The resulting spectrum generalizes the Moshinsky oscillator and, in
the fully coupled geometry, becomes implicit in the energy due to
the defect-renormalized angular index that depends on $\alpha$, $J_{t}$,
$J_{z}$, and the longitudinal momentum $k$.

By dissecting three salient configurations, balanced torsion ($J_{t}=J_{z}$),
purely spinning ($J_{z}=0$), and purely screw ($J_{t}=0$) alongside
the fully coupled case, we clarified the complementary roles of time-like
and space-like torsion: the former induces an $E$-dependent, self-consistent
shift of the angular index, whereas the latter introduces an explicit,
$k$ controlled bias. Together with the conical deficit ($\alpha<1$),
these ingredients broke flat-space degeneracies, skew $\ell$-resolved
level spacings, and couple longitudinal transport to the angular barrier. 

Geometric consistency further constrains the spectrum: positivity
of $g_{\phi\phi}$ enforces a minimal radius when $|J_{t}|>|J_{z}|$,
effectively imposing a hard-wall boundary that enhances the centrifugal
barrier---especially at small $|\ell|$ and shifts levels upward;
when $J_{t}=0$ or $|J_{t}|\le|J_{z}|$, the full radial range remains
accessible. These constraints are fully compatible with the confluent-hypergeometric
solvability and the quantization prescription employed across all
cases.

All limits continuously recover known results: switching off $J_{t}$
and/or $J_{z}$, or restoring $\alpha\to1$, yields the corresponding
sub-cases and, ultimately, the standard Moshinsky spectrum. Beyond
their formal interest, these findings illuminate spin--geometry--momentum
coupling in torsion-rich backgrounds and suggest concrete analog routes---graphene
with disclinations, trapped-ion and photonic platform for probing
the predicted defect-induced spectral signatures without ultra-high
energies.

\bibliographystyle{ChemEurJ}
\bibliography{refboumali}

\end{document}